\documentclass[prb,aps,twocolumn,superscriptaddress]{revtex4}

\usepackage{latexsym}
\usepackage{url}
\usepackage{color}
\usepackage{verbatim}
\usepackage{multirow}
\usepackage{amsmath}
\usepackage{mathrsfs}
\usepackage{float}
\usepackage[dvipsnames]{xcolor}
\usepackage{mathtools}
\usepackage{slashed}
\usepackage{physics}	% installed packages for typing maths and physics
\usepackage{graphicx}   % need for Fig.ures
\usepackage{epstopdf}
\usepackage{subfigure}  % use for side-by-side Fig.ures
\usepackage{hyperref}   % use for hypertext links, including those to external documents and URLs
\usepackage{wasysym}

\usepackage[latin1]{inputenc}
\usepackage{tikz}
\usetikzlibrary{decorations.pathmorphing}
\usetikzlibrary{shapes.misc}
\tikzset{cross/.style={cross out, draw=black, minimum size=8*(#1-\pgflinewidth), inner sep=0pt, outer sep=0pt},
default radius will be 1pt. 
cross/.default={1pt}}
\newcommand\numberthis{\addtocounter{equation}{1}\tag{\theequation}}

\begin{document}
\title{Superconductivity with repulsion: a variational approach}
\author{Laura Fanfarillo}
\affiliation{Istituto dei Sistemi Complessi (ISC-CNR), Via dei Taurini 19, I-00185 Rome, Italy}
\author{Yifu Cao}
\affiliation{Department of Physics, University of Florida, Gainesville, Florida, USA}
\author{Chandan Setty}
\affiliation{Department of Physics and Astronomy, Iowa State University, Ames, Iowa 50011, USA}
\affiliation{Ames National Laboratory, U.S. Department of Energy, Ames, Iowa 50011, USA}
\author{Sergio Caprara}
\affiliation{Dipartimento di Fisica, Sapienza Universit\`a di Roma, Piazzale Aldo Moro 5, I-00185 Rome, Italy}
\author{P. J. Hirschfeld}
\affiliation{Department of Physics, University of Florida, Gainesville, Florida, USA}

\begin{abstract}
We revisit the stability of the superconducting state within mean-field theory in the presence of repulsive pairing interactions, focusing on multiband systems where such channels naturally arise. We show that, when repulsion is present, the self-consistent BCS solution appears as a saddle point of the conventional mean-field free energy, casting doubt on its physical stability. We show that this pathology is an artifact of using a non-variational functional. Recasting the problem with Bogoliubov's variational principle restores a free energy that is bounded from below and places the BCS solution at a genuine minimum. Using a two-band toy model relevant to iron-based superconductors, we demonstrate the stability of the $s_\pm$ state and clarify how projection schemes that rely only on the interaction matrix can misidentify the attractive eigenmode that drives pairing. Our results clarify the instability issue highlighted by Aase \textit{et~al.} and provide a consistent foundation for analyzing fluctuations in the presence of repulsive interactions.
\end{abstract}

\maketitle
\section{Introduction}
A common feature of many recently discovered unconventional superconductors, from iron-based compounds to kagome materials, is the presence of multiple conduction bands and non-trivial intra- and interband pairing interactions. In multiband superconductors, the superconducting (SC) state arises from the coherent interplay of several condensates, whose mutual interference can be constructive or destructive depending on the interaction structure. This multicomponent character leads to novel phenomena, such as emergent collective excitations (e.g., Leggett modes\cite{Leggett_PTP1966, Blumberg_PRL2007, Klein_PRB2010, Murotani_PRB2017, Giorgianni_NatPhys2019,Nagashima_PRR2024}) and exotic phases breaking time-reversal symmetry\cite{Stanev_PRB2010, Garaud_PRL2011, Maiti_PRB2013, Marciani_PRB2013, Poniatowski_CommPhys2022}, among others.

The first theoretical extension of BCS theory to multiband systems was developed by Suhl, Matthias, and Walker\cite{Suhl_PRL1959} and Moskalenko \cite{Moskalenko_FMM1959}, who studied a two-band system with both intra- and interband interactions. Their analysis revealed that interband pairing can significantly alter the critical temperature and gap structure. Generalizing to $N$ bands, the pairing interaction becomes a matrix, and the BCS equations define $N$ coupled gaps. Unlike in the single-band case, a nontrivial SC solution can emerge even in the presence of both attractive and repulsive pairing channels, where the attractive/repulsive character is defined by the positive/negative eigenvalues of the interaction matrix. A well-known example is the $s\pm$ superconductivity in Fe-based materials which arises from strong interband pairing in systems with two or more bands\cite{Mazin_PRL2008, Kuroki_PRL2008, Chubukov_PRB2008, Fernandes_Nature2022}. 

Beyond the mean-field description, the interplay between attractive and repulsive pairing channels also has profound effects on SC fluctuations. Triggered by the discovery of new multiband superconductors, extensive studies have analyzed SC fluctuations in systems where both attractive and repulsive channels are present. In particular, it has been shown that the presence of repulsive channels qualitatively modifies the nature of SC fluctuations both above and below $T_c$ \cite{Fanfarillo_PRB2009, Marciani_PRB2013, Dalal_PRB2023}, with observable consequences in experiments\cite{Fanfarillo_SST2014, Cea_PRB2016, Yerin_PRB2017, Fiore_PRB2022, Nguyen_EPJ2023}. 

Recently, the mean-field framework has been revisited in the context of the stability of the BCS solution in the presence of repulsive pairing channels. A recent work by Aase {\it et al.}~[\onlinecite{Aase_PRB2023}] pointed out that the mean-field free energy becomes unbounded from below in the presence of repulsive eigenvalues. In such cases, the BCS solution corresponds to a saddle point, and fluctuation corrections appear ill-defined, since the system can escape along the unstable directions. To address this, Aase {\it et al.} proposed a dimensional reduction of the order-parameter space, retaining only the subspace associated with the attractive eigenvalues of the interaction. However, this kind of projection is physically unjustified, as there is no underlying symmetry or dominant interaction that would naturally restrict the space of SC order parameters.

In this work, we revisit the mean-field theory for multiband superconductors. We first show that the identification of attractive and repulsive pairing modes must be based on the full pairing kernel, not just on the interaction matrix. We then derive the free energy variationally and show that the resulting functional is always bounded from below, even in the presence of repulsive channels, thus guaranteeing the stability of the BCS solution without requiring a dimensional reduction of the order-parameter space. 

Our analysis builds on and generalizes a previous result by Agra {\it et al.}~[\onlinecite{Agra_EJP2006}], who showed, in the context of the antiferromagnetic Ising model, that the standard mean-field free energy, $F_\text{MF}$, does not necessarily possess a variational structure, and that a consistent variational formulation restores stability.
In a similar spirit, we clarify why $F_\text{MF}$ appears unstable in the presence of repulsive channels and show that this instability arises from treating $F_\text{MF}$ as a variational functional, when in fact it is not. The variational free energy not only reproduces the BCS gap equations via a saddle point condition, but also ensures a proper characterization of the energy landscape. Crucially, the analysis applies to \emph{any} multichannel decomposition, regardless of what the channels represent e.g. bands, orbitals, or momentum patches. Our results resolve the issue raised by Aase {\it et al.}, and provide a consistent framework for analyzing superconductivity in multichannel systems even in the presence of repulsion.

\section{Mean-field theory}
\label{Sec:MF}

\subsection{Model Hamiltonian and BCS Equations}

We consider a generic multiband BCS model for superconductivity with both intra/interband pair-hopping interactions. The Hamiltonian reads
\begin{equation}
H=\sum_{i,\mathbf{k},\sigma}\epsilon_{i\mathbf{k}} c^\dagger_{i\mathbf{k}\sigma}c_{i\mathbf{k}\sigma} - 
\sum_{\mathbf{k},\mathbf{k}'} {\vec{\phi}}_{\mathbf{k}}^{\ \dagger} \  \hat{V} \ \vec{\phi}_{\mathbf{k'}}.
\label{H}
\end{equation}
Here $i=1,...,N$ is the band index, $\epsilon_{i\mathbf{k}}$ is the $i$-th band dispersion and $c^{(\dagger)}_{i \mathbf{k} \sigma}$ destroys (creates) an electron in band $i$, with momentum $\mathbf{k}$ and spin $\sigma$. We use a matrix notation for the interacting Hamiltonian. The pairing operator is the $N$-dimensional vector $\phi^{\dagger}_{\mathbf{k}} = (\phi^{\dagger}_{1 \mathbf{k}} \ \ ... \ \ \phi^{\dagger}_{N \mathbf{k}})$ with $\phi^{\dagger}_{i \mathbf{k}} = c^{\dagger}_{i \mathbf{k} \uparrow}c^{\dagger}_{i -\mathbf{k}\downarrow}$. The effective interaction is a matrix in the band space, $\hat{V}$, whose elements, in the spirit of conventional BCS theory, are assumed to be finite and constant in a narrow energy window around the Fermi level. We follow the convention where attractive interactions correspond to $V_{ij}>0$.
The mean field Hamiltonian corresponding to Eq. \eqref{H} can be written as 
\begin{equation}
H_\text{MF} = \sum_{i, \mathbf{k}} \psi^{\dagger}_{i, \mathbf{k}} A_{i, \mathbf{k}}  \psi_{i, \mathbf{k}} + \sum_{i, \mathbf{k}}  \epsilon_{i\mathbf{k}} - \vec{\Delta}^* \hat{V}^{-1} \vec{\Delta}
\label{H_MF}
\end{equation}
where we define  $\Delta_i= \sum_j V_{ij} \langle \phi_{i \mathbf{k}}\rangle$ and introduce the vector $\vec{\Delta}^*= (\Delta^*_1...\Delta^*_N)$. We also use the Nambu notation with $\psi^{\dagger}_{i, \mathbf{k}} = ( c^\dagger_{i\mathbf{k}\uparrow} \ \ c_{i\mathbf{k}\downarrow})$ and 
\begin{equation}
\label{Aik}
A_{i, \mathbf{k}} =\left(
\begin{array}{cc}
\epsilon_{i, \mathbf{k}} & - \Delta_i
\\
- \Delta^*_i & -\epsilon_{i, \mathbf{k}}
\end{array}
\right).
\end{equation}
The partition function $Z_\text{MF}=\int D\psi \, D\psi^\dagger e^{-\beta H_\text{MF}}$, with $\beta=1/T$ and $T$ the temperature, can be computed explicitly within mean field since $H_\text{MF}$ is quadratic in the fermionic variable. The free energy $F_\text{MF}= - T \ln Z_{MF}$ reads 
\begin{align*}
    F_\text{MF}=&  \vec{\Delta}^* {\hat{V}^{-1}} \vec{\Delta}
    + \sum_{i, \mathbf{k}}(\epsilon_{i\mathbf{k}} - E_{i\mathbf{k}}) \\
    &-2T \sum_{i, \mathbf{k}} \ln{(1+e^{-\beta E_{i\mathbf{k}}})} \numberthis \label{F_MF},
\end{align*}
where $\pm E_{i\mathbf{k}}=\pm \sqrt{\epsilon_{i\mathbf{k}}^2+\lvert\Delta_i\rvert^2}$ is the quasi-particle spectrum defined by the eigenvalues of $A_{i, \mathbf{k}}$, Eq.(\ref{Aik}). 

The saddle point equation for the mean field free energy leads to the BCS gap equations
\begin{equation}
\frac{\partial F_\text{MF}}{\partial \vec{\Delta}^*}=0  \quad \quad \Rightarrow \quad \quad (\hat{V}^{-1} - \hat{\Pi} ) \vec{\Delta}= 0.
\label{gapeq}
\end{equation}
where $\hat{\Pi}=\text{diag}(\Pi_i)$ with $\Pi_i=\sum_{\mathbf{k}} \tanh{(\beta E_{i\mathbf{k}}/2)}/2E_{i\mathbf{k}}>0$ is the particle-particle bubble.

From Eq.~(\ref{gapeq}) we immediately observe that in a multiband superconductor, the BCS gap equations are in general not diagonal in band space. The particle-particle bubble $\hat{\Pi}$ is always diagonal in the band index, since each element $\Pi_i$ depends solely on the dispersion of the corresponding band $i$. In contrast, the interaction matrix $\hat{V}$ generally includes interband couplings and is therefore not diagonal. As a result, the band gaps $\Delta_i$ are typically not eigenmodes of the kernel, and one must diagonalize the full matrix $\hat{V}\hat{\Pi}$ to identify the true collective pairing channels that can be either attractive or repulsive. 
A controlled procedure for identifying these channels in the presence of repulsion has been discussed in the context of the analysis of fluctuations \cite{Fanfarillo_PRB2009, Marciani_PRB2013}. In this approach, the interaction matrix is first diagonalized, followed by a transformation that preserves the structure of the pairing interaction and diagonalizes $\hat{\Pi}$. In this way, the sign of each pairing channel is preserved and a consistent classification of modes as attractive or repulsive is ensured. This approach provides a consistent basis for identifying attractive and repulsive channels.
%even when $\hat{V}$ and $\hat{\Pi}$  do not commute. 
A rigorous and systematic treatment of the transformation used to define the eigenmodes in the general case goes beyond the scope of the present work and will be addressed in a future study~[\onlinecite{future_work}].

\subsection{Apparent instability of $F_{\rm MF}$}

As seen from Eq.(\ref{F_MF}), the behavior of the mean-field free energy at large $\Delta$ is dominated by the first term, proportional to $\vec{\Delta}^* \hat{V}^{-1} \vec{\Delta}$. 
If $\hat{V}^{-1}$ has at least one negative eigenvalue (corresponding to a repulsive pairing channel), then this term becomes unbounded from below along the associated direction. As a result, the free energy $F_\text{MF}$ has no global minimum, and the BCS solution represents a saddle point rather than a true minimum. This observation was recently emphasized by Aase {\it et al.} 
[\onlinecite{Aase_PRB2023}], who argued that fluctuation corrections around such saddle points are ill-defined due to the presence of unstable directions in the free energy landscape. 

To address this issue, Aase {\it et al.} proposed to regularize the theory by projecting out the repulsive sector, effectively restricting the theory to the subspace spanned by the attractive eigenmodes of $\hat{V}$. While this procedure removes directions associated with negative eigenvalues of $\hat{V}$, it does not, in general, eliminate the unbounded directions of the full free energy functional. This is because, as we discussed above, the physical instability directions are determined by the full kernel $\hat{V}\hat{\Pi}$, whose eigenmodes coincide with those of $\hat{V}$ only in the special case where $\hat{\Pi}\propto \hat{I}$, i.e., when all bands are identical. When the bands differ, a projection based solely on the interaction matrix fails to remove unbounded directions, as we will demonstrate explicitly in the two-band model discussed below.

\section{Variational Free-energy }

The apparent instability of the BCS solution in the presence of repulsive pairing channels is resolved by adopting a fully variational formulation, as discussed, for example, in Ref.~[\onlinecite{Agra_EJP2006}]. We begin by rewriting the Hamiltonian as $H=H_\text{MF} + (H-H_\text{MF})$  so that the exact free-energy takes the form $F=F_\text{MF} - T\ln  \langle e^{- \beta (H-H_\text{MF})}\rangle_{\text{MF}}$, where the angular brackets denote the thermal average taken with respect to the mean-field Gibbs weight, $\exp(-\beta H_\text{MF})$ with $\beta=1/T$. Using the convexity of the exponential function, we obtain an upper bound for the free energy 
\begin{equation}
F \le F_\text{MF} + \langle H-H_\text{MF} \rangle_{\text{MF}} = F_{\text{Var}}
\label{F_Var}
\end{equation}
The above, also known as Bogoliubov inequality, provides a practical route to approximating the true free energy. By minimizing $F_{\text{Var}}$ with respect to the trial parameters - in our case, the set of band gaps $\{\Delta_i\}$ - one obtains the best approximation to the full free energy $F$ within the mean-field framework.

Notice that the argument rests solely on the Bogoliubov variational inequality and on the fact that the trial Hamiltonian $H_{\mathrm{MF}}$, Eq.~(\ref{H_MF}), is quadratic in the Nambu spinors~$\Psi$, and is therefore exactly solvable. It places no restriction on what the channel index $i = 1,\dots,N$ represents.  Consequently, the same cancellation that removes the term $\vec{\Delta}^{*}\hat V^{-1}\vec{\Delta}$ operates in \emph{any} orthonormal decomposition of the pairing field, whether the
components label electronic bands, orbitals, momentum patches, or real-space clusters. Whenever the interaction matrix in a given basis possesses a negative eigenvalue, the mean-field free energy is unbounded along that direction, while the variational correction $\langle H - H_{\mathrm{MF}}\rangle_{\mathrm{MF}}$ restores a free
energy that is bounded from below.

Below, we illustrate the behavior of $F_{\text{MF}}$ and $F_\text{Var}$ in two pedagogical examples: (i) a single-band model with attractive/repulsive interaction, and (ii) a two-band model with interband interaction only. As we will see, the correction term $\langle H-H_\text{MF} \rangle_{\text{MF}}$, which quantifies the energetic "distance" between the trial mean-field Hamiltonian and the full Hamiltonian, modifies $F_\text{MF}$ only quantitatively when the interaction is attractive. In contrast, when repulsive channels are present, this correction cures the pathological large-$\Delta$ behavior of  $F_\text{MF}$, ensuring a well-defined free energy landscape $F_{\text{Var}}$ even in directions associated with repulsion.

Contrary to   Ref.~[\onlinecite{Aase_PRB2023}], the variational mean-field procedure naturally yields a bounded and physically meaningful functional $F_{\text{Var}}$, without requiring any artificial reduction of the order-parameter space. In particular, as we will show explicitly in the following examples, the dimensionality of the problem is preserved, and the stability of the SC state can be analyzed directly within the full parameter space.

\subsection{Example 1: Single-Band Superconductor}

Let us consider a single-band system characterized by an effective interaction $V$, which can be either attractive ($V>0$) or repulsive ($V<0$). While the discussion holds regardless of the specific electronic dispersion, for concreteness we consider a tight-binding model on a square lattice.

Following the steps discussed in the previous section we define a mean-field free-energy $F_\text{MF}$ as given in Eq.(\ref{F_MF}) with $i=1$ as we have only one band. 
The expression in Eq.(\ref{F_MF}) is valid for both attractive and repulsive cases and the only term that depends explicitly on the sign of $V$ is the first one, $V^{-1}|\Delta|^2$. 

\begin{figure}[tb]
\centering
\includegraphics[width=0.99\linewidth]{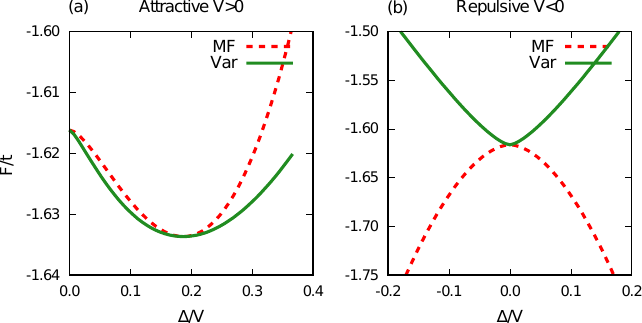}
\caption{Energy landscape of a single-band superconductor versus the dimensionless gap $\Delta/V$. 
Panels (a) and (b) display the attractive ($V>0$) and repulsive ($V<0$) case, respectively. $F_{\text{MF}}$, dashed line and $F_{\text{Var}}$, solid line, both renormalized to the hopping $t$. The self-consistent BCS gap $\bar{\Delta}$ satisfies the saddle point condition for both functionals. In (a) $\bar{\Delta}$ sits at a true minimum of both $F_{\text{MF}}$ and $F_{\text{Var}}$ with the variational correction being only quantitative. In (b) $F_{\text{MF}}$ is unbounded from below, while  $F_{\text{Var}}$ remains bounded and attains its global minimum at $\bar{\Delta}=0$. Parameters $t=1$ eV, $V=\pm 2 t$, $T=0$.}
\label{1band_FMF}
\end{figure}

The saddle point condition determines the BCS gap equation $(V^{-1} - \Pi)\Delta=0$. As expected the BCS equation admits a non-trivial solution for the attractive case only, while in the repulsive case $\bar{\Delta}=0$ at any temperature. 
The behavior of $F_\text{MF}$ as a function of $\Delta$ around the saddle point for both cases is shown in Fig.(\ref{1band_FMF}). The first term in Eq.(\ref{F_MF}), $V^{-1}|\Delta|^2$, changes sign depending on the sign of the interaction $V$ and leads to an $F_\text{MF}$ that is unbounded below in the repulsive case.  

\begin{figure*}[tb]
\centering
\includegraphics[width=0.98\textwidth]{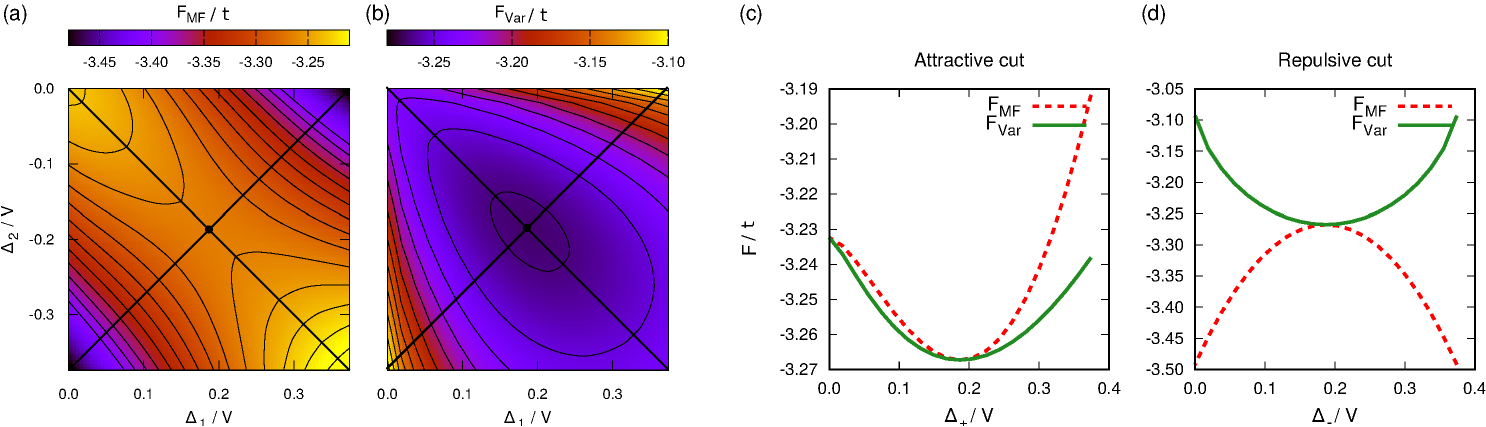}
\caption{Free energy landscape for a two-band model with purely interband repulsion, $V_{12}=V_{21}=V<0$.  
(a) Color map of $F_{\rm MF}/t$ in terms of $(\Delta_{1}/V ,\Delta_{2}/V)$ from Eq.~\eqref{F_MF}.  
(b) Variational free energy $F_{\rm Var}/t$, obtained from Eqs.~\eqref{F_Var} and \eqref{spm_Hav}. Black diagonal lines show the attractive and repulsive fluctuating modes around the BCS solution as defined by the diagonalization of the full kernel. $F_{\rm MF}$ is unbounded from below along the repulsive direction, placing the BCS $s_{\pm}$ solution at a saddle point, $F_{\rm Var}$ instead remains bounded and exhibits a true minimum at the BCS solution.  (c,d) Value of $F_{\rm MF}$ (dashed) and $F_{\rm Var}$ (solid) along the attractive and repulsive directions, respectively. These cuts mirror the single-band behavior of Fig.~\ref{1band_FMF}. For the attractive mode the variational correction is only quantitative, whereas along the repulsive mode it cures the unboundedness of $F_{\rm MF}$ and restores a stable minimum.  
Parameters $t_{1}=t_{2}=1$ eV, $V_{12}=V_{21}=V= -2 t$, $T=0$.}
\label{spm_FMF}
\end{figure*}

To compute $F_{\text{Var}}$ we need to evaluate $\langle H-H_\text{MF} \rangle_{\text{MF}}$, i.e.,
\begin{equation}
\bigg \langle  - V \sum_{\mathbf{k},\mathbf{k}'}  \phi^\dagger_{\mathbf{k}} \, \phi_{\mathbf{k}'} + \sum_{\mathbf{k}} ( \Delta \phi^\dagger_{\mathbf{k}} + \Delta^* \phi_{\mathbf{k}})  - V^{-1}|\Delta|^2   \bigg \rangle_{\text{MF}} 
\label{1band_Hav}
\end{equation} 
from which one sees already that the last term of Eq.(\ref{1band_Hav}), once added to $F_\text{MF}$, Eq.(\ref{F_MF}), will cancel the term $V^{-1}|\Delta|^2$ responsible for pathological behavior of $F_\text{MF}$. Since $H_\text{MF}$ is quadratic, we can perform the average in Eq.(\ref{1band_Hav}) using the Wick theorem and evaluating the single terms $\langle \phi^{(\dagger)}_{\mathbf{k}} \rangle_\text{MF} = \Delta^{(*)} \tanh{(\beta E_{\mathbf{k}}/2)}/2E_{\mathbf{k}}$. The final result can be written in a compact form as 
\begin{equation}
\langle H-H_\text{MF} \rangle_{\text{MF}} = - V|\Delta|^2 \big ( V^{-1} - \Pi \big )^2 .
\end{equation}
This correction is zero at the saddle point, where $V^{-1} - \Pi = 0$, but finite elsewhere.
The behavior of $F_\text{Var}$ as a function of $\Delta$ around the saddle point for both attractive and repulsive cases is shown in Fig.(\ref{1band_FMF}).
When $F_\text{Var}$ is minimized, $\Delta$ takes the value that solves the BCS gap equation, i.e, it has the same saddle point of $F_{\text{MF}}$. However, now the BCS solution is always a minimum of $F_\text{Var}$. In particular, thanks to the cancellation of $V^{-1}|\Delta|^2$, the variational free energy is bounded from below even for $V<0$.

In this pedagogical example, we showed that even in the absence of attractive pairing, the solution $\Delta=0$, which is the only stationary point of the functional, is correctly identified as a stable minimum of $F_\text{Var}$. Notice that, in this case, the dimensional reduction proposed in \cite{Aase_PRB2023}, eliminating the repulsive pairing channel from the order parameter phase-space would leave the stability issue undetermined, constraining the free-energy to the single $\Delta=0$ point. 

\subsection{Example 2: Two-Band $s_{\pm}$ Superconductor}

The variational approach discussed for the single-band model can be easily extended to the multiband case. As a minimal example, we consider a two-band system in which one repulsive pairing channel is present, i.e., $\det \hat{V}<0$. This situation can arise either in systems with dominant intraband pairing and repulsion in one band, or in systems where interband interactions dominate. We focus on the last example and, without loss of generality, we consider the limiting scenario with $V_{11/22}=0$, $V_{12/21}=V < 0$, a simple yet widely used toy model for Fe-based superconductors. As we will briefly comment at the end of this section, the same analysis and conclusions apply to the case of dominant attractive interband interaction, $V > 0$.

For a two-band system with interband interaction only, the saddle point condition for $F_\text{MF}$, defined in Eq.(\ref{F_MF}), yields a set of coupled equations for the  gaps found by solving Eqs.(\ref{gapeq}), which explicitly read
\begin{equation}
\label{2band BCS}
    \begin{cases}
      \Delta_2/V - \Delta_1 \Pi_1 =  0  \\
      \Delta_1/V - \Delta_2 \Pi_2 =  0   
    \end{cases}
\end{equation}
Unlike the single-band case, these equations admit a nontrivial solution even when $V<0$, thanks to the additional degree of freedom provided by the second band. In particular, a consistent solution requires opposite signs of the gaps, $\text{sign} \, \bar{\Delta}_1= - \text{sign}\,\bar{\Delta}_2$, corresponding to the so-called $s\pm$ state typical of many Fe-based superconductors\cite{Fernandes_Nature2022}.
The non-diagonal structure of Eq.~(\ref{2band BCS}) reflects the fact that the band gaps $\Delta_i$ are not eigenmodes of the kernel. As discussed in Sec.~\ref{Sec:MF}, the eigenmodes are defined by the diagonalization of $\hat{V}\hat{\Pi}$, which decouples the BCS equations. In the special case of equivalent bands, where $\hat{\Pi}\propto \hat{I}$, this reduces to diagonalizing the interaction matrix. The resulting eigenvalues are $\lambda_{A/R} = \pm|V|$, with corresponding attractive and repulsive eigenmodes $\Delta_{A/R} = (\Delta_1 \mp \Delta_2)/\sqrt{2}$. For simplicity, we begin by considering this symmetric case,  and will return to the general case of two inequivalent bands later to demonstrate the limitations of the projection proposed in Ref.~[\onlinecite{Aase_PRB2023}].\\

We now examine the behavior of $F_\text{MF} (\Delta_1, \Delta_2)$, in the vicinity of the $s_\pm$ solution $\bar{\Delta}_1=-\bar{\Delta}_2$. While the outcomes of the analysis do not depend on the specific band structure, for concreteness we consider a two-band tight-binding model on a square lattice and assume $t_1=t_2$. As shown in Fig.~\ref{spm_FMF}(a), the presence of the repulsive eigenvalue $\lambda_{R}$ causes $F_\text{MF}$ to be unbounded from below along the direction $\Delta_{R}$ associated with the repulsive mode. 

To compute $F_{\text{Var}}$, we evaluate the correction  $\langle H-H_\text{MF} \rangle_{\text{MF}}$,  which in this case reads
\begin{equation}
\bigg \langle  \bigg( - V \sum_{\mathbf{k},\mathbf{k}'} \phi^\dagger_{1 \mathbf{k}} \, \phi_{2\mathbf{k}'} + \sum_{ i \mathbf{k}} \Delta_i \phi^\dagger_{i \mathbf{k}}  - V^{-1} \Delta_1^* \Delta_2  \bigg) + h.c.\bigg \rangle_{\text{MF}} 
\label{spm_Hav}
\end{equation} 
As in the single-band case, the term $V^{-1} (\Delta_1^* \Delta_2 + \Delta_1^* \Delta_2)$ cancels the repulsive contribution in $F_\text{MF}$ that causes the pathological behavior at large $\Delta_i$. The calculation proceeds as a straightforward generalization of the single-band case. 
The resulting variational free energy is shown in Fig.(\ref{spm_FMF})(b). The BCS $s_{\pm}$ solutions now correspond to true minima of $F_{\text{Var}}$, which is bounded from below across the entire order-parameter space, including along the direction of the repulsive eigenmode $\Delta_R$.
To highlight the analogy with the single-band case, in panels (c) and (d) we plot the values of $F_{\text{MF}}$ and $F_{\text{Var}}$ along the attractive and repulsive eigenmode directions, respectively. As seen previously in Fig.~\ref{1band_FMF}, the variational correction modifies the free energy only quantitatively along the attractive direction, while it qualitatively reshapes the landscape along the repulsive one, curing the unboundedness of $F_{\text{MF}}$ . \\

\begin{figure}[tb]
\centering
\includegraphics[width=\linewidth]{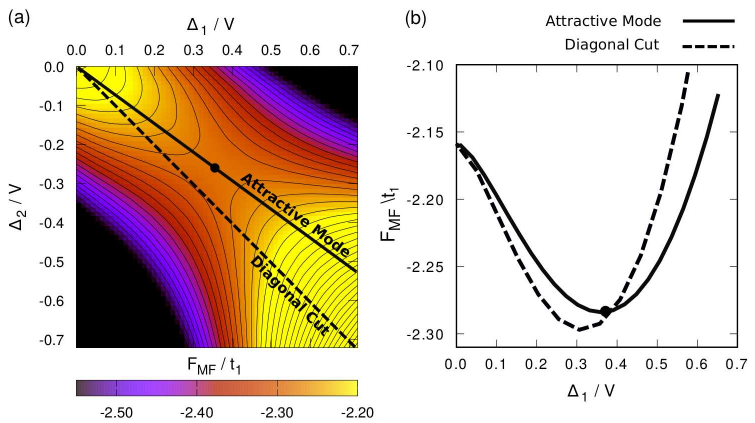}
\caption{Identification of the attractive mode in the two-band system with inequivalent bands and purely interband repulsion. (a) Color map of $F_{\text{MF}}(\Delta_1, \Delta_2)$ normalized to $t_1$, showing the $s_\pm$ solution. The solid line, $\Delta_1 = -\alpha, \Delta_2$, with $\alpha = \sqrt{\Pi_2/\Pi_1}$, corresponds to the attractive eigenmode of the full kernel $\hat{V}\hat{\Pi}$, the critical mode defining the SC instability. The dashed line, $\Delta_1 = -\Delta_2$, corresponds to the "attractive" eigenmode of the interaction matrix $\hat{V}$. This subspace does not include the $s_\pm$ solution and does not align to the attractive mode direction. Even requiring the diagonal projection to pass through the saddle-point solution would still fail to identify the instability line. As discussed in the text, the two attractive subspaces coincide only when the bands are equivalent. (b) Cuts of $F_{\text{MF}}/t_1$ along the two directions. The diagonalization of the full BCS kernel identifies the true attractive mode, while the projection based solely on $\hat{V}$ does not contain the BCS solution. Parameters $t_1=3t_2=1$ eV, $V = - 2 t_1$, $T = 0$.}
\label{projection}
\end{figure}

%Inequivalent band - identification of the critical mode
We now turn to the more general case of two inequivalent bands. Since all steps of the analysis proceed identically to the symmetric case, we do not repeat them here. In particular, we still find that the self-consistent BCS solution corresponds to a saddle point of the mean-field free energy $F_{\text{MF}}$, while the variational free energy $F_{\text{Var}}$ remains well-defined and yields a true minimum. The only quantitative difference is the location of the BCS solution: due to the asymmetry in the band structure, it no longer lies at $\bar{\Delta}_1 = -\bar{\Delta}_2$, but rather at $\bar{\Delta}_1 = -\alpha \bar{\Delta}_2$, with $\alpha = \sqrt{\Pi_2/\Pi_1}$.

As previously anticipated, this asymmetric case is particularly instructive, as it sharpens the distinction between the true instability direction and the attractive subspace identified by diagonalizing the interaction matrix alone. Since the interaction matrix is unchanged, $V_{11} = V_{22} = 0$, $V_{12} = V_{21} = V < 0$, it is diagonalized by the symmetric and antisymmetric combinations of the band gaps, $\Delta_{\mp} = (\Delta_1 \mp \Delta_2)/\sqrt{2}$
which correspond to the attractive and repulsive eigenmodes of $\hat{V}$. However, because $\hat{\Pi}$ is no longer proportional to the identity, these directions do not diagonalize the full kernel $\hat{V}\hat{\Pi}$. The true eigenmodes, which define the physical collective pairing channels, become asymmetric combinantion, $\Delta_{A/R} = (\alpha \Delta_1 \mp \alpha^{-1} \Delta_2)/\sqrt{2}$, with $ \alpha = \sqrt{\Pi_2/\Pi_1}$. 
This explicitly shows that for inequivalent bands, the projection method proposed in Ref.~[\onlinecite{Aase_PRB2023}] is not aligned with the actual fluctuation eigenmode.

% \st{It therefore fails to isolate the correct instability direction and mixes attractive and repulsive components.}

In Fig.~\ref{projection}, we illustrate this issue in a specific case with $t_1 = 3t_2$. Panel (a) shows the mean-field free energy $F_{\text{MF}}(\Delta_1, \Delta_2)$, with two lines indicating the attractive eigenmode of the full kernel $\hat{V}\hat{\Pi}$ (solid line), and the "attractive" subspace obtained from diagonalizing $\hat{V}$ alone (dashed line). Panel (b) shows $F_{\text{MF}}$ along these cuts. The eigenmode of $\hat{V}\hat{\Pi}$ defines the correct critical mode: it crosses the $s_\pm$ solution and identify the attractive fluctuations subspace around it. In contrast, the direction based on $\hat{V}$ alone not only misses the BCS solution, but also mixes attractive and repulsive components. Notice that, even forcing the diagonal projection to cross the $s_\pm$ solution, it would be still not aligned with the instability direction retaining contributions from both attractive and repulsive channels.\\

%2band with dominant attractive interaction (s++ case) 
As a final remark, we recall that the same conclusions apply when the dominant interband interaction is attractive. Even when $V_{12} = V_{21} = V > 0$, the interaction matrix still satisfies $\det\hat{V} < 0$, and the mean-field free energy $F_{\text{MF}}(\Delta_1, \Delta_2)$ remains unbounded along the direction associated with the repulsive eigenmode of $\hat{V}\hat{\Pi}$. As a result, the BCS $s_{++}$ solution also corresponds to a saddle point of $F_\text{MF}$ rather than a minimum. As before, the variational correction $\langle H - H_\text{MF} \rangle_{\text{MF}}$ removes the unbounded behavior and restores a well-defined free-energy minimum at the BCS solution, just as in the $s_\pm$ case.

\section{Conclusions}

In contrast to the results reported in Ref.~[\onlinecite{Aase_PRB2023}], where a stability problem was identified based on an analysis of the mean-field free energy $F_\text{MF}$, our work demonstrates that this apparent issue arises from the use of $F_\text{MF}$, which is not a variational functional, to assess stability.
By adopting a variational approach, we derived a bounded free energy, $F_{\text{Var}} = F_\text{MF} + \langle H-H_\text{MF} \rangle_{\text{MF}}$, that correctly reproduces the BCS gap equations and ensures stability even in the presence of repulsive pairing channels.
Our detailed analysis of both a single-band and a two-band model explicitly demonstrates that the variational correction cures the pathological behavior of $F_\text{MF}$ without the need for any artificial dimensional reduction of the order parameter space.

We further clarify that the identification of attractive and repulsive pairing modes must be based on the full kernel $\hat{V}\hat{\Pi}$, not on the interaction matrix alone. Projections based solely on $\hat{V}^{-1}$ do not identify the true instability direction and mix repulsive and attractive channels, potentially mischaracterizing the nature of fluctuations.

The variational framework thus provides a robust and physically consistent method for analyzing the stability and fluctuation properties of superconductors with competing pairing interactions. Remarkably, the variational approach is completely basis-agnostic: the channel index $i$ may label bands, orbital, or momentum patches; the same variational cure applies to \emph{all} multichannel
superconductors. The approach can also be generalized to analyze the stability in Eliashberg theory in which repulsion exists in certain frequency domains \cite{Dalal_PRB2023}. 

Our results resolve an important conceptual ambiguity in the treatment of repulsive channels and lay the foundation for future investigations into fluctuation effects and collective modes in complex multiband superconductors.

\section{Acknowledgements} We thank Rafael Fernandes, Jonathan Ruhman and Niels Henrik Aase for useful discussions. C.S. acknowledges support from Iowa State University and Ames National Laboratory startup funds.  S.C. acknowledges support from the University of Rome Sapienza, projects Ateneo 2023 (RM123188E830D258) and Ateneo 2024 (RM124190C54BE48D). Y.C. and P.H. were supported in part by NSF-DMR-2231821. Part of the research was performed at the Kavli Institute for Theoretical Physics of Santa Barbara supported by the National Science Foundation under Grant Nos. NSF PHY-1748958 and NSF PHY-2309135.\\ \newline
\noindent

\bibliographystyle{apsrev4-1}%Choose a bibliograhpic style
\bibliography{stability.bib}
\end{document}